\documentclass[review]{elsarticle}
\biboptions{sort&compress}
\usepackage{lineno,hyperref}
\usepackage{threeparttable}
\usepackage{multirow}
\usepackage{longtable,booktabs}
\usepackage[]{caption2}

\modulolinenumbers[5]

\journal{Journal of Quantitative Spectroscopy and Radiative Transfer}

\begin{document}

\begin{frontmatter}

\title{Theoretical investigation on the soft X-ray spectrum of the highly-charged W$^{54+}$ ions}


\author[nwnu]{Xiaobin Ding\corref{mycorrespondingauthor}}
\cortext[mycorrespondingauthor]{Corresponding author.}
\ead{dingxb@nwnu.edu.cn}
\author[nwnu]{Jiaoxia Yang}
\author[sophiaU]{Fumihiro Koike}
\author[NIFS]{Izumi Murakami}
\author[NIFS]{Daiji Kato}
\author[NIFS]{Hiroyuki A Sakaue}
\author[UEC]{Nobuyuki Nakamura}
\author[nwnu]{Chenzhong Dong}

\address[nwnu]{Key Laboratory of Atomic and Molecular Physics and Functional Materials
of Gansu Province, College of Physics and Electronic Engineering, Northwest Normal University, Lanzhou 730070, China}
\address[sophiaU]{Department of physics, Sophia University, Tokyo 102-8554, Japan}
\address[NIFS]{National Institute for Fusion Science, Toki, Gifu 509-5292, Japan}
\address[UEC]{Institute for Laser Science, The University of Electro-Communications, Chofu, Tokyo 182-8585, Japan}

\begin{abstract}
A detailed level collisional-radiative model of the E1 transition spectrum of Ca-like W$^{54+}$ ion has been constructed. All the necessary atomic data has been calculated by relativistic configuration interaction (RCI) method with the implementation of Flexible Atomic Code (FAC). The results are in reasonable agreement with the available experimental and previous theoretical data. The synthetic spectrum has explained the EBIT spectrum in 29.5-32.5 \AA \,, while several new strong transitions has been proposed to be observed in 18.5-19.6 \AA \, for the future EBIT experiment with electron density $n_e$ = $10^{12}$ cm$^{-3}$ and electron beam energy $E_e$ = 18.2 keV.
\end{abstract}


\end{frontmatter}

%\linenumbers

\section{Introduction}

Tungsten (W) and its alloy had been used as the armor plate material for the plasma facing component in the divertor region of ITER (International Thermonuclear Experimental Reactor Tokamak) and the other magnetic confinement fusion reactors because of their high melting point, low sputtering yield, and low tritium retention rate\cite{0029-5515-45-3-007,Matthews2009934}. However, tungsten impurity ions might be produced during the plasma-wall interaction in the edge and then they might be transported to the high-temperature core plasma region. In the core plasma, these tungsten impurities can be ionized further to highly charged ions and radiate high energy photons. Therefore, the large radiation loss can be caused by these highly charged tungsten impurity ions, which will lead to the plasma disruption if the relative concentration of W ions in the core plasma is higher than about 10$^{-5}$\cite{Radtke2001}. Monitoring and controlling the flux of the highly charged W impurity ions are crucial to the success of the fusion\cite{1402-4896-2009-T134-014022}. Thus a thorough knowledge of the atomic properties of tungsten ions were strongly needed by the magnetic confinement fusion research. Furthermore, the spectra of tungsten impurity ions observed from the fusion plasma also provide plenty of important information about the fusion plasma parameters such as electron density, electron temperature, and ion temperature. Therefore, it can also be used to diagnose the plasma.

There are many research works related to the energy levels and transition properties of tungsten in various ionization stage in the past several decades\cite{Fei2014,Kramida2006,
 Kramida2009,Kramida2011,20072_S1060,0953-4075-45-3-035003}, where only a few studies have focused on W$^{54+}$ ion\cite{0953-4075-50-4-045004,Dipti2015Electron,PhysRevA.87.062505,Ralchenko2008, 0953-4075-43-7-074026,0953-4075-44-19-195007,PhysRevA.83.032517}. Y. Ralchenko \emph{et al.} used the electron beam ion trap (EBIT) to observe the M1 spectrum from 3d$^n$(n=1-9) ground state fine structure multiplets of Co-like W$^{47+}$ to K-like W$^{55+}$ ions and a non-Maxwellian collisional-radiative model (CRM) was used to analyze the observed spectrum\cite{PhysRevA.83.032517}. U. I. Safronova and A. S. Safronova calculated the wavelength and transition rates of the magnetic dipole (M1) and electric quadrupole (E2) transitions between the multiplets of the ground state configuration ([Ne]3s$^2$3p$^6$3d$^2$) of W$^{54+}$ ion by using relativistic many-body perturbation theory (RMBPT)\cite{0953-4075-43-7-074026}. P. Quinet used a full relativistic Dirac-Fock method to calculate the line wavelengthes and transition rates of the forbidden transitions within the 3p$^k$(k=1-5) and 3d$^n$(n=1-9) ground state configuration multiplets from Al-like W$^{61+}$ to Co-like W$^{47+}$ ions\cite{0953-4075-44-19-195007}. Dipti \emph{et al.} calculated the excitation energies and the electron collisional excitation cross sections of Ca-like W$^{54+}$ ion by relativistic distorted wave theory\cite{Dipti2015Electron}. C. F. Fischer \emph{et al.} found the core correlation is very important for the transition energies of 3d$^k$ configuration in tungsten ions\cite{Atomcs}.

 T. Lennartsson \emph{et al.} observed the electric dipole (E1) transitions from the excited states [Ne]3s$^2$3p$^5$3d$^3$ to the ground state [Ne]3s$^2$3p$^6$3d$^2$ of W$^{54+}$ ion in the wavelength range of 26.3-43.5\AA\, in the Lawrence Livermore National Laboratory (LLNL) EBIT at the electron beam energy of 18.2 keV, and a CRM was applied to explain the observed spectrum\cite{PhysRevA.87.062505}. Ding and his collaborators have performed a calculation on the E1, E2, M1, M2 transitions between [Ne]3s$^2$3p$^5$3d$^3$ and [Ne]3s$^2$3p$^6$3d$^2$ levels of W$^{54+}$ ion by using an MCDF method with the restricted active space method; the relativistic and electron correlation effects as well as the Breit interaction and some Quantum Electrodynamic effects have been taken into account\cite{0953-4075-50-4-045004,1701-05504,Ding2017}. The results are in reasonable agreement with available experimental data for both M1 and E1 transition lines. Several strong E1 transitions have been predicted. Some of these strong transitions are in good agreement with the observation from the EBIT, while others were not observed even in the similar wavelength range with similar transition rates. And some transitions with wavelength in 18.5-19.6 \AA\, are suggested to be observed in the future experiment.

  The present paper focuses on the explanation of the E1 transition spectrum of W$^{54+}$ ion from the EBIT with the electron density $n_e = 10^{12}$ cm$^{-3}$ and the energy of electron beam $E_e$ = 18.2 keV by CRM. The present paper is constructed as follows. In section 2, the theory of CRM and the calculation of the necessary atomic data are described. The result of the present calculation and the discussion are given in Section 3. Finally, the conclusion is presented in section 4.

\section{Theoretical method}

The CRM is one of the most useful simulation methods for the spectrum from the optically thin and isotropic plasma. It has been widely used to study the spectrum of highly charged ions in X-ray, VUV and visible region\cite{0953-4075-48-14-144028,Ding2012Collisional,0953-4075-47-17-175002}. In order to carry out the analysis of the fine structure of the spectrum, a detailed-level CRM should be used\cite{Ding2016Collisional,Ding2012Collisional,0953-4075-48-14-144029,0953-4075-47-17-175002}.

 The spectral intensity $I_{i,j}( \lambda)$ of a transition from the upper level $i$ to the lower level $j$ with wavelength $\lambda$ can be defined as:
\begin{eqnarray}\label{eq1}
   I_{i,j}(\lambda) \propto n(i) A(i,j)\phi(\lambda).
\end{eqnarray}
 where $A(i,j)$ is the transition rate for the transition from the energy level $i$ to $j$, which can be obtained by experimental observations or by accurate theoretical calculations. The function $\phi(\lambda)$ is the normalized line profile, which was taken as a Gaussian profile to include the effect of Doppler, natural, collisional and instrumental broadenings in the present work. The notation $n(i)$ is the population of the ions in the excited upper level $i$, which was determined by the detailed atomic physics processes in the plasma, e.g., spontaneous radiative transitions, collisional excitation and deexcitation, collisional ionization, radiative recombination and three-body recombination etc.. These atomic processes can be calculated by using an appropriate theory.

 For the plasma in the EBIT, which is in low electron density and the energy distribution of the free electron is almost mono-energy, the radiative recombination, three-body recombination, electron collisional ionization, and dielectronic recombination processes are expected to be negligible in this situation. Only the electron collisional excitations, deexcitations, and spontaneous radiative transitions processes will determine the population $n(i)$ of the excited upper level $i$. For a specifically excited level $i$, the temporal development of the population $n(i)$ can be obtained by solving the collisional-radiative rate equations:
 \begin{eqnarray}
 \label{eq2}\nonumber
 \frac{{\rm d}}{{\rm d} t}n(i)&=\sum_{j<i}C(j,i)n_en(j) \\
                             &-[{\sum_{j<i}F(i,j)n_e+A(i,j)}+\sum_{j>i}C(i,j)n_e]n(i)\\\nonumber
                             &+\sum_{j>i}[F(j,i)n_e+A(j,i)]n(j).
 \end{eqnarray}
 where ${n_e}$ is the electron density of the plasma, $C(i,j)$ and $F(j,i)$ are collisional excitation and deexcitation rates coefficient from the level $j$ to $i$, respectively. These rate coefficient can be calculated by convoluting the cross section of the collisional (de) excitation processes with the free electron energy distribution function, which can be assumed as Maxwellian distribution, for the plasma in local thermodynamic equilibrium (LTE) with the electron temperature $T_e$. However, the electron energy distribution in the electron beam of the EBIT is more like mono-energy distribution function instead of Maxwellian distribution. Thus, the rate coefficient for the atomic processes in the EBIT may be calculated by taking the electron energy distribution as the $\delta$ function. These rate equations are solved under the Quasi-Steady-State (QSS) approximation, i.e., $ {\rm d}n(i)/{{\rm d} t} = 0$. Finally, the intensity $I_{i,j}(\lambda)$ of the specific transition can be calculated when the population $n(i)$ of the upper level $i$ of the transition is obtained.

 Because the W$^{54+}$ ion is a heavy highly charged ion, the relativistic effects will play an important role in the energy level structure and transition properties. The ground state of W$^{54+}$ ion is [Ne]3s$^{2}$3p$^6$3d$^2$, which have two electrons in the open $3d$ subshells. Therefore, the relativistic and electron correlation effects should both be taken into account in the theoretical calculation. The present calculation is performed by using the relativistic configuration interaction (RCI) method with the implementation of FAC packages\cite{Gu2008The}. The atomic data including the energy levels, radiative transition rates, and cross sections of collisional (de) excitation are calculated. The collisional deexcitation process is the inverse of collisional excitation process. Therefore, the collisional deexcitation rate coefficient can be calculated by the principle of detailed balance from the corresponding electron collisional excitation process. In the present calculation, most of the important configuration interaction are included by single and double electron substitution from n=3 shells to n=4 subshells (nSD). For instance, the configuration interaction from configuration 3s$^2$3p$^6$3dnl(n=4), 3s$^2$3p$^5$3d$^2$nl (n=3, 4), 3s$^1$3p$^6$3d$^2$nl (n=3, 4), 3s$^2$3p$^6$nln'l' (n, n'=4), 3s$^2$3p$^4$3d$^2$nln'l' (n, n'=3, 4), 3p$^6$3d$^2$nln'l' (n, n'=3, 4), 3s$^2$3p$^5$3dnln'l' (n, n'=3, 4), 3s$^1$3p$^6$3dnln'l' (n, n'=3, 4) were taken into account in the present calculation. The E1 radiative transitions, electron collisional excitation processees between the levels of the configuration 3p$^6$3d$^2$, 3p$^5$3d$^3$, 3p$^6$3d4l are included to simulate the spectrum of W$^{54+}$ ion.

\section{Result and discussion}

The level energies (in eV) of the ground state 3p$^6$3d$^2$ and the first excited state 3p$^5$3d$^3$ multiplets of W$^{54+}$ ion were presented in Table~\ref{Tab1}. The levels are sorted by their excitation energies. There are 9 levels in the ground state 3p$^6$3d$^2$ and 110 levels in the first excited state 3p$^5$3d$^3$. The $jj$ coupling labels are provided to designate the levels. The level energies calculated from different electron correlation models (nSD) are given to show the configuration interaction effects on the excitation energies. It can be inferred from the table that the level energies converge along with the increase of configurations interaction. The result of the present calculation agrees well with previous MCDF calculations\cite{Ding2017}. The discrepancy was found to be about 0.12\%. For the levels of ground configuration, the results from the NIST database\cite{nist} were compared with the present calculation. The discrepancy was about 0.19\%. A good agreement between the present result and previous result was found indicating that the most of the important configuration interaction effects has been included in the present work.

\scriptsize
\LTright=3pt
\LTleft=3pt
\begin{longtable}{ccccccc}
\caption{The level energies (in eV) of the ground state 3p$^6$3d$^2$ and the first excited state 3p$^5$3d$^3$ of W$^{54+}$ ion. The column DF is the level energy calculated with Dirac-Hartree-Fock approximation. The fourth and fifth column labeled as cal(3SD) and cal(4SD) represent the level energy obtained by considering the configuration interaction from single and double substitution in n=3 and n=4 subshells.
The column 'Theo.' and 'NIST' represent the previous available data obtained by MCDF calcualtion and the NIST database\cite{nist} for comparison.\label{Tab1}}\\
 \hline \hline\noalign{\vskip3pt}
 Index   &Levels   & DF    & cal(3SD)    & cal(4SD)    & Theo.$^a$ & NIST    \\[3pt]
  \hline\noalign{\vskip3pt}
\endfirsthead
 \multicolumn{6}{@{}c@{}}{Table  \ref{Tab1} Continued\dots}\\
 \hline \hline\noalign{\vskip3pt}
 Index   &Levels   & DF    & cal(3SD)    & cal(4SD)    & Theo.$^a$ & NIST    \\[3pt]
  \hline\noalign{\vskip3pt}
\endhead
 \hline
 \noalign{\vskip3pt}
 \multicolumn{6}{r}{\itshape Continued\dots}\\
\endfoot
 \noalign{\vskip3pt}\hline
\endlastfoot
  $Ground state 3s^{2}3p^{6}3d^{2}$ \\\hline
1	&	[3p$^{6}$3d$_{3/2}^{2}$]$_{2}$	&	0	&	0 	&	0	&	0	&	0\\	
2	&	[3p$^{6}$3d$_{3/2}^{2}$]$_{0}$	&	24.317	&	24.173 	&	23.248	&	23.123	&	23.309\\	
3	&	[3p$^{6}$3d$_{3/2}$3d$_{5/2}$]$_{3}$	&	71.992	&	72.023 	&	72.321	&	72.456	&	72.59\\	
4	&	[3p$^{6}$3d$_{3/2}$3d$_{5/2}$]$_{2}$	&	83.069	&	83.067 	&	82.736	&	82.805	&	82.882\\	
5	&	[3p$^{6}$3d$_{3/2}$3d$_{5/2}$]$_{4}$	&	86.125	&	86.118 	&	86.300	&	86.359	&	86.4\\	
6	&	[3p$^{6}$3d$_{3/2}$3d$_{5/2}$]$_{1}$	&	88.161	&	88.138 	&	87.597	&	87.613	&	87.962\\	
7	&	[3p$^{6}$3d$_{5/2}^{2}$]$_{4}$	&	152.557	&	152.596 	&	152.933	&	153.158	&	153\\	
8	&	[3p$^{6}$3d$_{5/2}^{2}$]$_{2}$	&	161.371	&	161.384 	&	160.966	&	161.160	&	161.1\\	
9	&	[3p$^{6}$3d$_{5/2}^{2}$]$_{0}$	&	186.754	&	186.482 	&	185.041	&	185.140	&	185.5\\	\hline
  $Excited state 3s^{2}3p^{5}3d^{3}$\\\hline
10	&	[(3p$_{1/2}^{2}$3p$_{3/2}^{3}$)$_{3/2}$(3d$_{5/2}^{3}$)$_{3/2}$]$_{2}$	&	271.864	&	267.172 	&	273.656	&	273.769	&	\\
11	&	[(3p$_{1/2}^{2}$3p$_{3/2}^{3}$)$_{3/2}$(3d$_{3/2}^{3}$)$_{3/2}$]$_{1}$	&	276.776	&	272.177 	&	278.597	&	278.693	&	\\
12	&	[(3p$_{1/2}^{2}$3p$_{3/2}^{3}$)$_{3/2}$(3d$_{3/2}^{3}$)$_{3/2}$]$_{0}$	&	281.019	&	276.325 	&	282.713	&	282.774	&	\\
13	&	[(3p$_{1/2}^{2}$3p$_{3/2}^{3}$)$_{3/2}$(3d$_{5/2}^{3}$)$_{3/2}$]$_{3}$	&	282.739	&	278.073 	&	284.507	&	284.582	&	\\
14	&	[((3p$_{1/2}^{2}$3p$_{3/2}^{3}$)$_{3/2}$(3d$_{3/2}^{2}$)$_{2}$)$_{3/2}$3d$_{5/2}$]$_{3}$	&	334.793	&	330.688 	&	337.025	&	337.262	&	\\
15	&	[((3p$_{1/2}^{2}$3p$_{3/2}^{3}$)$_{3/2}$(3d$_{3/2}^{2}$)$_{2}$)$_{5/2}$3d$_{5/2}$]$_{4}$	&	340.543	&	336.665 	&	342.952	&	343.159	&	\\
16	&	[((3p$_{1/2}^{2}$3p$_{3/2}^{3}$)$_{3/2}$(3d$_{3/2}^{2}$)$_{2}$)$_{3/2}$3d$_{5/2}$]$_{2}$	&	342.013	&	337.558 	&	343.879	&	344.043	&	\\
17	&	[((3p$_{1/2}^{2}$3p$_{3/2}^{3}$)$_{3/2}$(3d$_{3/2}^{2}$)$_{2}$)$_{1/2}$3d$_{5/2}$]$_{2}$	&	344.868	&	340.627 	&	346.909	&	347.135	&	\\
18	&	[((3p$_{1/2}^{2}$3p$_{3/2}^{3}$)$_{3/2}$(3d$_{3/2}^{2}$)$_{2}$)$_{3/2}$3d$_{5/2}$]$_{1}$	&	346.876	&	342.447 	&	348.711	&	348.901	&	\\
19	&	[((3p$_{1/2}^{2}$3p$_{3/2}^{3}$)$_{3/2}$(3d$_{3/2}^{2}$)$_{2}$)$_{5/2}$3d$_{5/2}$]$_{5}$	&	348.958	&	345.129 	&	351.385	&	351.541	&	\\
20	&	[((3p$_{1/2}^{2}$3p$_{3/2}^{3}$)$_{3/2}$(3d$_{3/2}^{2}$)$_{2}$)$_{7/2}$3d$_{5/2}$]$_{3}$	&	349.878	&	345.632 	&	351.879	&	352.048	&	\\
21	&	[((3p$_{1/2}^{2}$3p$_{3/2}^{3}$)$_{3/2}$(3d$_{3/2}^{2}$)$_{2}$)$_{5/2}$3d$_{5/2}$]$_{0}$	&	349.968	&	345.632 	&	351.926	&	352.089	&	\\
22	&	[((3p$_{1/2}^{2}$3p$_{3/2}^{3}$)$_{3/2}$(3d$_{3/2}^{2}$)$_{2}$)$_{7/2}$3d$_{5/2}$]$_{6}$	&	350.377	&	346.507 	&	352.741	&	352.878	&	\\
23	&	[((3p$_{1/2}^{2}$3p$_{3/2}^{3}$)$_{3/2}$(3d$_{3/2}^{2}$)$_{2}$)$_{3/2}$3d$_{5/2}$]$_{4}$	&	350.991	&	346.859 	&	353.116	&	353.284	&	\\
24	&	[((3p$_{1/2}^{2}$3p$_{3/2}^{3}$)$_{3/2}$(3d$_{3/2}^{2}$)$_{2}$)$_{7/2}$3d$_{5/2}$]$_{4}$	&	357.312	&	353.222 	&	359.492	&	359.671	&	\\
25	&	[((3p$_{1/2}^{2}$3p$_{3/2}^{3}$)$_{3/2}$(3d$_{3/2}^{2}$)$_{2}$)$_{5/2}$3d$_{5/2}$]$_{2}$	&	359.202	&	354.601 	&	360.928	&	361.129	&	\\
26	&	[((3p$_{1/2}^{2}$3p$_{3/2}^{3}$)$_{3/2}$(3d$_{3/2}^{2}$)$_{2}$)$_{5/2}$3d$_{5/2}$]$_{3}$	&	359.647	&	355.507 	&	361.754	&	361.952	&	\\
27	&	[((3p$_{1/2}^{2}$3p$_{3/2}^{3}$)$_{3/2}$(3d$_{3/2}^{2}$)$_{2}$)$_{5/2}$3d$_{5/2}$]$_{1}$	&	359.997	&	355.537 	&	361.840	&	362.029	&	\\
28	&	[((3p$_{1/2}^{2}$3p$_{3/2}^{3}$)$_{3/2}$(3d$_{3/2}^{2}$)$_{2}$)$_{7/2}$3d$_{5/2}$]$_{5}$	&	365.560	&	361.917 	&	368.102	&	368.256	&	\\
29	&	[((3p$_{1/2}^{2}$3p$_{3/2}^{3}$)$_{3/2}$(3d$_{3/2}^{2}$)$_{0}$)$_{3/2}$3d$_{5/2}$]$_{4}$	&	377.203	&	372.566 	&	378.857	&	378.929	&	\\
30	&	[((3p$_{1/2}^{2}$3p$_{3/2}^{3}$)$_{3/2}$(3d$_{3/2}^{2}$)$_{0}$)$_{3/2}$3d$_{5/2}$]$_{2}$	&	381.494	&	376.402 	&	382.637	&	382.657	&	\\
31	&	[((3p$_{1/2}^{2}$3p$_{3/2}^{3}$)$_{3/2}$(3d$_{3/2}^{2}$)$_{2}$)$_{7/2}$3d$_{5/2}$]$_{1}$	&	390.190	&	383.853 	&	390.051	&	390.056	&	\\
32	&	[((3p$_{1/2}^{2}$3p$_{3/2}^{3}$)$_{3/2}$(3d$_{3/2}^{2}$)$_{0}$)$_{3/2}$3d$_{5/2}$]$_{3}$	&	390.357	&	384.849 	&	390.889	&	390.731	&	\\
33	&	[((3p$_{1/2}^{2}$3p$_{3/2}^{3}$)$_{3/2}$(3d$_{3/2}^{2}$)$_{2}$)$_{5/2}$3d$_{5/2}$]$_{2}$	&	393.665	&	387.333 	&	393.325	&	393.157	&	\\
34	&	[((3p$_{1/2}^{2}$3p$_{3/2}^{3}$)$_{3/2}$(3d$_{3/2}^{2}$)$_{0}$)$_{3/2}$3d$_{5/2}$]$_{3}$	&	395.220	&	388.874 	&	394.857	&	394.733	&	\\
35	&	[((3p$_{1/2}^{2}$3p$_{3/2}^{3}$)$_{3/2}$3d$_{3/2}$)$_{0}$(3d$_{5/2}^{2}$)$_{4}$]$_{4}$	&	410.421	&	406.464 	&	412.697	&	413.038	&	\\
36	&	[((3p$_{1/2}^{2}$3p$_{3/2}^{3}$)$_{3/2}$3d$_{3/2}$)$_{1}$(3d$_{5/2}^{2}$)$_{4}$]$_{5}$	&	414.800	&	411.268 	&	417.424	&	417.736	&	\\
37	&	[((3p$_{1/2}^{2}$3p$_{3/2}^{3}$)$_{3/2}$3d$_{3/2}$)$_{0}$(3d$_{5/2}^{2}$)$_{2}$]$_{2}$	&	416.667	&	412.128 	&	418.373	&	418.689	&	\\
38	&	[((3p$_{1/2}^{2}$3p$_{3/2}^{3}$)$_{3/2}$(3d$_{3/2}^{2}$)$_{0}$)$_{3/2}$3d$_{5/2}$]$_{1}$	&	418.451	&	414.824 	&	420.451	&	419.866	&	\\
39	&	[((3p$_{1/2}^{2}$3p$_{3/2}^{3}$)$_{3/2}$3d$_{3/2}$)$_{3}$(3d$_{5/2}^{2}$)$_{4}$]$_{6}$	&	420.957	&	415.297 	&	421.437	&	421.736	&	\\
40	&	[((3p$_{1/2}^{2}$3p$_{3/2}^{3}$)$_{3/2}$3d$_{3/2}$)$_{1}$(3d$_{5/2}^{2}$)$_{4}$]$_{3}$	&	425.424	&	421.069 	&	427.235	&	427.500	&	\\
41	&	[((3p$_{1/2}^{2}$3p$_{3/2}^{3}$)$_{3/2}$3d$_{3/2}$)$_{1}$(3d$_{5/2}^{2}$)$_{2}$]$_{2}$	&	425.944	&	422.356 	&	428.549	&	428.802	&	\\
42	&	[((3p$_{1/2}^{2}$3p$_{3/2}^{3}$)$_{3/2}$3d$_{3/2}$)$_{3}$(3d$_{5/2}^{2}$)$_{4}$]$_{5}$	&	425.991	&	422.467 	&	428.574	&	428.881	&	\\
43	&	[((3p$_{1/2}^{2}$3p$_{3/2}^{3}$)$_{3/2}$3d$_{3/2}$)$_{1}$(3d$_{5/2}^{2}$)$_{2}$]$_{1}$	&	426.923	&	422.600 	&	428.819	&	429.044	&	\\
44	&	[((3p$_{1/2}^{2}$3p$_{3/2}^{3}$)$_{3/2}$3d$_{3/2}$)$_{3}$(3d$_{5/2}^{2}$)$_{4}$]$_{7}$	&	427.623	&	423.133 	&	429.239	&	429.467	&	\\
45	&	[((3p$_{1/2}^{2}$3p$_{3/2}^{3}$)$_{3/2}$3d$_{3/2}$)$_{1}$(3d$_{5/2}^{2}$)$_{4}$]$_{4}$	&	427.672	&	423.334 	&	429.534	&	429.789	&	\\
46	&	[((3p$_{1/2}^{2}$3p$_{3/2}^{3}$)$_{3/2}$3d$_{3/2}$)$_{1}$(3d$_{5/2}^{2}$)$_{2}$]$_{3}$	&	432.445	&	427.718 	&	433.929	&	434.203	&	\\
47	&	[((3p$_{1/2}^{2}$3p$_{3/2}^{3}$)$_{3/2}$3d$_{3/2}$)$_{2}$(3d$_{5/2}^{2}$)$_{4}$]$_{4}$	&	437.602	&	433.178 	&	439.202	&	439.326	&	\\
48	&	[((3p$_{1/2}^{2}$3p$_{3/2}^{3}$)$_{3/2}$3d$_{3/2}$)$_{2}$(3d$_{5/2}^{2}$)$_{4}$]$_{3}$	&	438.543	&	434.006 	&	440.144	&	440.334	&	\\
49	&	[((3p$_{1/2}^{2}$3p$_{3/2}^{3}$)$_{3/2}$3d$_{3/2}$)$_{3}$(3d$_{5/2}^{2}$)$_{2}$]$_{1}$	&	438.641	&	434.784 	&	441.037	&	441.338	&	\\
50	&	[((3p$_{1/2}^{2}$3p$_{3/2}^{3}$)$_{3/2}$3d$_{3/2}$)$_{2}$(3d$_{5/2}^{2}$)$_{4}$]$_{6}$	&	439.931	&	435.500 	&	441.552	&	441.773	&	\\
51	&	[((3p$_{1/2}^{2}$3p$_{3/2}^{3}$)$_{3/2}$3d$_{3/2}$)$_{3}$(3d$_{5/2}^{2}$)$_{2}$]$_{4}$	&	440.325	&	435.762 	&	441.869	&	442.007	&	\\
52	&	[((3p$_{1/2}^{2}$3p$_{3/2}^{3}$)$_{3/2}$3d$_{3/2}$)$_{3}$(3d$_{5/2}^{2}$)$_{2}$]$_{5}$	&	440.336	&	436.009 	&	442.079	&	442.165	&	\\
53	&	[((3p$_{1/2}^{2}$3p$_{3/2}^{3}$)$_{3/2}$3d$_{3/2}$)$_{0}$(3d$_{5/2}^{2}$)$_{0}$]$_{0}$	&	444.847	&	439.450 	&	445.855	&	446.163	&	\\
54	&	[((3p$_{1/2}^{2}$3p$_{3/2}^{3}$)$_{3/2}$3d$_{3/2}$)$_{2}$(3d$_{5/2}^{2}$)$_{2}$]$_{2}$	&	447.059	&	442.116 	&	448.282	&	448.481	&	\\
55	&	[((3p$_{1/2}^{2}$3p$_{3/2}^{3}$)$_{3/2}$3d$_{3/2}$)$_{3}$(3d$_{5/2}^{2}$)$_{4}$]$_{3}$	&	452.440	&	447.295 	&	453.385	&	453.580	&	\\
56	&	[((3p$_{1/2}^{2}$3p$_{3/2}^{3}$)$_{3/2}$3d$_{3/2}$)$_{3}$(3d$_{5/2}^{2}$)$_{4}$]$_{4}$	&	456.750	&	451.980 	&	458.032	&	458.198	&	\\
57	&	[((3p$_{1/2}^{2}$3p$_{3/2}^{3}$)$_{3/2}$3d$_{3/2}$)$_{2}$(3d$_{5/2}^{2}$)$_{4}$]$_{2}$	&	460.988	&	455.504 	&	461.538	&	461.647	&	\\
58	&	[((3p$_{1/2}^{2}$3p$_{3/2}^{3}$)$_{3/2}$3d$_{3/2}$)$_{2}$(3d$_{5/2}^{2}$)$_{4}$]$_{5}$	&	464.235	&	459.539 	&	465.368	&	465.389	&	\\
59	&	[((3p$_{1/2}^{2}$3p$_{3/2}^{3}$)$_{3/2}$3d$_{3/2}$)$_{1}$(3d$_{5/2}^{2}$)$_{0}$]$_{1}$	&	465.621	&	460.007 	&	466.285	&	466.474	&	\\
60	&	[((3p$_{1/2}^{2}$3p$_{3/2}^{3}$)$_{3/2}$3d$_{3/2}$)$_{3}$(3d$_{5/2}^{2}$)$_{0}$]$_{3}$	&	470.088	&	464.555 	&	470.752	&	470.828	&	\\
61	&	[((3p$_{1/2}^{2}$3p$_{3/2}^{3}$)$_{3/2}$3d$_{3/2}$)$_{3}$(3d$_{5/2}^{2}$)$_{4}$]$_{2}$	&	471.224	&	466.182 	&	471.920	&	471.719	&	\\
62	&	[((3p$_{1/2}^{2}$3p$_{3/2}^{3}$)$_{3/2}$3d$_{3/2}$)$_{3}$(3d$_{5/2}^{2}$)$_{4}$]$_{1}$	&	474.370	&	467.146 	&	473.333	&	473.503	&	\\
63	&	[((3p$_{1/2}^{2}$3p$_{3/2}^{3}$)$_{3/2}$3d$_{3/2}$)$_{2}$(3d$_{5/2}^{2}$)$_{0}$]$_{2}$	&	474.387	&	468.779 	&	474.956	&	475.067	&	\\
64	&	[((3p$_{1/2}^{2}$3p$_{3/2}^{3}$)$_{3/2}$3d$_{3/2}$)$_{2}$(3d$_{5/2}^{2}$)$_{2}$]$_{3}$	&	476.453	&	469.685 	&	475.569	&	475.534	&	\\
65	&	[((3p$_{1/2}^{2}$3p$_{3/2}^{3}$)$_{3/2}$3d$_{3/2}$)$_{3}$(3d$_{5/2}^{2}$)$_{2}$]$_{4}$	&	477.385	&	470.400 	&	476.323	&	476.313	&	\\
66	&	[((3p$_{1/2}^{2}$3p$_{3/2}^{3}$)$_{3/2}$3d$_{3/2}$)$_{2}$(3d$_{5/2}^{2}$)$_{0}$]$_{2}$	&	481.909	&	475.078 	&	480.941	&	480.731	&	\\
67	&	[((3p$_{1/2}^{2}$3p$_{3/2}^{3}$)$_{3/2}$3d$_{3/2}$)$_{1}$(3d$_{5/2}^{2}$)$_{2}$]$_{1}$	&	486.173	&	479.787 	&	485.476	&	485.082	&	\\
68	&	[((3p$_{1/2}^{2}$3p$_{3/2}^{3}$)$_{3/2}$3d$_{3/2}$)$_{2}$(3d$_{5/2}^{2}$)$_{2}$]$_{0}$	&	486.665	&	481.009 	&	486.558	&	486.092	&	\\
69	&	[((3p$_{1/2}^{2}$3p$_{3/2}^{3}$)$_{3/2}$3d$_{3/2}$)$_{3}$(3d$_{5/2}^{2}$)$_{2}$]$_{3}$	&	487.142	&	482.322 	&	487.985	&	487.630	&	\\
70	&	[(3p$_{1/2}^{2}$3p$_{3/2}^{3}$)$_{3/2}$(3d$_{5/2}^{3}$)$_{9/2}$]$_{6}$	&	493.589	&	490.274 	&	496.363	&	496.802	&	\\
71	&	[(3p$_{1/2}^{2}$3p$_{3/2}^{3}$)$_{3/2}$(3d$_{5/2}^{3}$)$_{3/2}$]$_{0}$	&	504.879	&	500.470 	&	506.656	&	507.002	&	\\
72	&	[(3p$_{1/2}^{2}$3p$_{3/2}^{3}$)$_{3/2}$(3d$_{5/2}^{3}$)$_{9/2}$]$_{5}$	&	505.289	&	500.800 	&	506.733	&	507.038	&	\\
73	&	[(3p$_{1/2}^{2}$3p$_{3/2}^{3}$)$_{3/2}$(3d$_{5/2}^{3}$)$_{3/2}$]$_{3}$	&	506.786	&	502.182 	&	508.242	&	508.543	&	\\
74	&	[(3p$_{1/2}^{2}$3p$_{3/2}^{3}$)$_{3/2}$(3d$_{5/2}^{3}$)$_{3/2}$]$_{2}$	&	523.308	&	517.979 	&	524.093	&	524.365	&	\\
75	&	[(3p$_{1/2}^{2}$3p$_{3/2}^{3}$)$_{3/2}$(3d$_{5/2}^{3}$)$_{5/2}$]$_{4}$	&	524.286	&	519.340 	&	525.606	&	525.942	&	\\
76	&	[(3p$_{1/2}^{2}$3p$_{3/2}^{3}$)$_{3/2}$(3d$_{5/2}^{3}$)$_{5/2}$]$_{2}$	&	535.660	&	530.254 	&	536.143	&	536.137	&	\\
77	&	[(3p$_{1/2}^{2}$3p$_{3/2}^{3}$)$_{3/2}$(3d$_{5/2}^{3}$)$_{9/2}$]$_{4}$	&	540.624	&	534.476 	&	540.343	&	540.525	&	\\
78	&	[(3p$_{1/2}^{2}$3p$_{3/2}^{3}$)$_{3/2}$(3d$_{5/2}^{3}$)$_{5/2}$]$_{3}$	&	544.344	&	538.414 	&	544.495	&	544.801	&	\\
79	&	[(3p$_{1/2}^{2}$3p$_{3/2}^{3}$)$_{3/2}$(3d$_{5/2}^{3}$)$_{5/2}$]$_{3}$	&	553.354	&	548.107 	&	553.875	&	553.794	&	\\
80	&	[(3p$_{1/2}^{2}$3p$_{3/2}^{3}$)$_{3/2}$(3d$_{5/2}^{3}$)$_{3/2}$]$_{1}$	&	556.678	&	550.699 	&	556.634	&	556.773	&	\\
81	&	[(3p$_{1/2}^{2}$3p$_{3/2}^{3}$)$_{3/2}$(3d$_{5/2}^{3}$)$_{5/2}$]$_{1}$	&	567.604	&	558.998 	&	565.219	&	565.442	&	\\
82	&	[(3p$_{1/2}$3p$_{3/2}^{4}$)$_{1/2}$(3d$_{3/2}^{3}$)$_{3/2}$]$_{2}$	&	632.534	&	626.627 	&	632.604	&	632.492	&	\\
83	&	[(3p$_{1/2}$3p$_{3/2}^{4}$)$_{1/2}$(3d$_{3/2}^{3}$)$_{3/2}$]$_{1}$	&	669.418	&	660.920 	&	666.964	&	666.827	&	\\
84	&	[((3p$_{1/2}$3p$_{3/2}^{4}$)$_{1/2}$(3d$_{3/2}^{2}$)$_{2}$)$_{5/2}$3d$_{5/2}$]$_{3}$	&	685.896	&	680.855 	&	686.909	&	687.221	&	\\
85	&	[((3p$_{1/2}$3p$_{3/2}^{4}$)$_{1/2}$(3d$_{3/2}^{2}$)$_{2}$)$_{5/2}$3d$_{5/2}$]$_{2}$	&	686.471	&	681.280 	&	687.301	&	687.555	&	\\
86	&	[((3p$_{1/2}$3p$_{3/2}^{4}$)$_{1/2}$(3d$_{3/2}^{2}$)$_{2}$)$_{5/2}$3d$_{5/2}$]$_{1}$	&	691.022	&	685.843 	&	691.842	&	692.014	&	\\
87	&	[((3p$_{1/2}$3p$_{3/2}^{4}$)$_{1/2}$(3d$_{3/2}^{2}$)$_{2}$)$_{5/2}$3d$_{5/2}$]$_{4}$	&	692.811	&	687.884 	&	693.899	&	694.156	&	\\
88	&	[((3p$_{1/2}$3p$_{3/2}^{4}$)$_{1/2}$(3d$_{3/2}^{2}$)$_{2}$)$_{5/2}$3d$_{5/2}$]$_{0}$	&	693.116	&	688.045 	&	694.040	&	694.211	&	\\
89	&	[((3p$_{1/2}$3p$_{3/2}^{4}$)$_{1/2}$(3d$_{3/2}^{2}$)$_{2}$)$_{5/2}$3d$_{5/2}$]$_{5}$	&	694.625	&	690.205 	&	696.166	&	696.384	&	\\
90	&	[((3p$_{1/2}$3p$_{3/2}^{4}$)$_{1/2}$(3d$_{3/2}^{2}$)$_{0}$)$_{1/2}$3d$_{5/2}$]$_{2}$	&	726.627	&	720.330 	&	726.189	&	726.150	&	\\
91	&	[((3p$_{1/2}$3p$_{3/2}^{4}$)$_{1/2}$(3d$_{3/2}^{2}$)$_{2}$)$_{3/2}$3d$_{5/2}$]$_{3}$	&	731.048	&	724.068 	&	729.856	&	729.825	&	\\
92	&	[((3p$_{1/2}$3p$_{3/2}^{4}$)$_{1/2}$(3d$_{3/2}^{2}$)$_{2}$)$_{3/2}$3d$_{5/2}$]$_{4}$	&	738.006	&	730.188 	&	735.993	&	735.966	&	\\
93	&	[((3p$_{1/2}$3p$_{3/2}^{4}$)$_{1/2}$(3d$_{3/2}^{2}$)$_{2}$)$_{3/2}$3d$_{5/2}$]$_{2}$	&	739.656	&	732.355 	&	738.170	&	738.129	&	\\
94	&	[((3p$_{1/2}$3p$_{3/2}^{4}$)$_{1/2}$(3d$_{3/2}^{2}$)$_{2}$)$_{3/2}$3d$_{5/2}$]$_{1}$	&	743.215	&	734.282 	&	740.378	&	740.497	&	\\
95	&	[((3p$_{1/2}$3p$_{3/2}^{4}$)$_{1/2}$(3d$_{3/2}^{2}$)$_{0}$)$_{1/2}$3d$_{5/2}$]$_{3}$	&	744.170	&	738.029 	&	743.776	&	743.576	&	\\
96	&	[((3p$_{1/2}$3p$_{3/2}^{4}$)$_{1/2}$(3d$_{3/2}$)$_{3/2}$)$_{2}$(3d$_{5/2}^{2}$)$_{4}$]$_{4}$	&	752.738	&	748.314 	&	754.373	&	754.976	&	\\
97	&	[((3p$_{1/2}$3p$_{3/2}^{4}$)$_{1/2}$(3d$_{3/2}$)$_{3/2}$)$_{2}$(3d$_{5/2}^{2}$)$_{4}$]$_{3}$	&	756.827	&	751.744 	&	757.770	&	758.279	&	\\
98	&	[((3p$_{1/2}$3p$_{3/2}^{4}$)$_{1/2}$(3d$_{3/2}$)$_{3/2}$)$_{2}$(3d$_{5/2}^{2}$)$_{4}$]$_{5}$	&	759.325	&	755.182 	&	761.205	&	761.763	&	\\
99	&	[((3p$_{1/2}$3p$_{3/2}^{4}$)$_{1/2}$(3d$_{3/2}$)$_{3/2}$)$_{2}$(3d$_{5/2}^{2}$)$_{2}$]$_{0}$	&	763.576	&	758.015 	&	764.190	&	764.767	&	\\
100	&	[((3p$_{1/2}$3p$_{3/2}^{4}$)$_{1/2}$(3d$_{3/2}$)$_{3/2}$)$_{2}$(3d$_{5/2}^{2}$)$_{2}$]$_{1}$	&	763.984	&	758.436 	&	764.604	&	765.173	&	\\
101	&	[((3p$_{1/2}$3p$_{3/2}^{4}$)$_{1/2}$(3d$_{3/2}$)$_{3/2}$)$_{2}$(3d$_{5/2}^{2}$)$_{4}$]$_{2}$	&	764.909	&	759.220 	&	765.349	&	765.853	&	\\
102	&	[((3p$_{1/2}$3p$_{3/2}^{4}$)$_{1/2}$(3d$_{3/2}$)$_{3/2}$)$_{2}$(3d$_{5/2}^{2}$)$_{2}$]$_{2}$	&	766.354	&	760.652 	&	766.765	&	767.260	&	\\
103	&	[((3p$_{1/2}$3p$_{3/2}^{4}$)$_{1/2}$(3d$_{3/2}$)$_{3/2}$)$_{2}$(3d$_{5/2}^{2}$)$_{4}$]$_{6}$	&	769.565	&	765.786 	&	771.778	&	772.272	&	\\
104	&	[((3p$_{1/2}$3p$_{3/2}^{4}$)$_{1/2}$(3d$_{3/2}$)$_{3/2}$)$_{2}$(3d$_{5/2}^{2}$)$_{2}$]$_{4}$	&	773.439	&	768.403 	&	774.429	&	774.884	&	\\
105	&	[((3p$_{1/2}$3p$_{3/2}^{4}$)$_{1/2}$(3d$_{3/2}$)$_{3/2}$)$_{2}$(3d$_{5/2}^{2}$)$_{2}$]$_{3}$	&	776.986	&	771.433 	&	777.559	&	778.043	&	\\
106	&	[((3p$_{1/2}$3p$_{3/2}^{4}$)$_{1/2}$(3d$_{3/2}$)$_{3/2}$)$_{2}$(3d$_{5/2}^{2}$)$_{0}$]$_{2}$	&	796.183	&	789.842 	&	796.082	&	796.470	&	\\
107	&	[((3p$_{1/2}$3p$_{3/2}^{4}$)$_{1/2}$(3d$_{3/2}$)$_{3/2}$)$_{1}$(3d$_{5/2}^{2}$)$_{4}$]$_{5}$	&	797.381	&	791.151 	&	796.687	&	796.760	&	\\
108	&	[((3p$_{1/2}$3p$_{3/2}^{4}$)$_{1/2}$(3d$_{3/2}$)$_{3/2}$)$_{1}$(3d$_{5/2}^{2}$)$_{4}$]$_{4}$	&	798.948	&	792.159 	&	797.999	&	798.247	&	\\
109	&	[((3p$_{1/2}$3p$_{3/2}^{4}$)$_{1/2}$(3d$_{3/2}$)$_{3/2}$)$_{1}$(3d$_{5/2}^{2}$)$_{2}$]$_{3}$	&	803.296	&	796.136 	&	802.009	&	802.234	&	\\
110	&	[((3p$_{1/2}$3p$_{3/2}^{4}$)$_{1/2}$(3d$_{3/2}$)$_{3/2}$)$_{1}$(3d$_{5/2}^{2}$)$_{2}$]$_{1}$	&	806.282	&	798.409 	&	804.420	&	804.706	&	\\
111	&	[((3p$_{1/2}$3p$_{3/2}^{4}$)$_{1/2}$(3d$_{3/2}$)$_{3/2}$)$_{1}$(3d$_{5/2}^{2}$)$_{4}$]$_{3}$	&	811.530	&	805.282 	&	810.977	&	810.831	&	\\
112	&	[((3p$_{1/2}$3p$_{3/2}^{4}$)$_{1/2}$(3d$_{3/2}$)$_{3/2}$)$_{1}$(3d$_{5/2}^{2}$)$_{2}$]$_{2}$	&	812.190	&	805.501 	&	811.002	&	811.046	&	\\
113	&	[((3p$_{1/2}$3p$_{3/2}^{4}$)$_{1/2}$(3d$_{3/2}$)$_{3/2}$)$_{1}$(3d$_{5/2}^{2}$)$_{0}$]$_{1}$	&	831.396	&	823.862 	&	829.747	&	829.814	&	\\
114	&	[(3p$_{1/2}$3p$_{3/2}^{4}$)$_{1/2}$(3d$_{5/2}^{3}$)$_{9/2}$]$_{4}$	&	831.475	&	826.499 	&	832.469	&	833.226	&	\\
115	&	[(3p$_{1/2}$3p$_{3/2}^{4}$)$_{1/2}$(3d$_{5/2}^{3}$)$_{9/2}$]$_{5}$	&	840.088	&	835.570 	&	841.539	&	842.263	&	\\
116	&	[(3p$_{1/2}$3p$_{3/2}^{4}$)$_{1/2}$(3d$_{5/2}^{3}$)$_{3/2}$]$_{1}$	&	851.857	&	845.181 	&	851.236	&	851.798	&	\\
117	&	[(3p$_{1/2}$3p$_{3/2}^{4}$)$_{1/2}$(3d$_{5/2}^{3}$)$_{3/2}$]$_{2}$	&	852.971	&	847.109 	&	853.197	&	853.855	&	\\
118	&	[(3p$_{1/2}$3p$_{3/2}^{4}$)$_{1/2}$(3d$_{5/2}^{3}$)$_{5/2}$]$_{2}$	&	859.833	&	853.175 	&	859.378	&	860.010	&	\\
119	&	[(3p$_{1/2}$3p$_{3/2}^{4}$)$_{1/2}$(3d$_{5/2}^{3}$)$_{5/2}$]$_{3}$	&	864.667	&	858.388 	&	864.592	&	865.208	&	\\

\end{longtable}
\begin{tablenotes}
\item[] $^{\rm a}$From Ding et al by MCDF method\cite{Ding2017}.
\end{tablenotes}
\normalsize

 According to the present calculation, the wavelength of E1 transition from the first excited state 3s$^2$3p$^5$3d$^3$ to the ground configuration state 3s$^2$3p$^6$3d$^2$ covered the range of 18.5-32.5 \AA. The strong transition with large transition rates was concentrated in two wavelength range 29.5-32.5 \AA\, and 18.5-19.6 \AA.

 The transition wavelength $\lambda$ (in \AA), transition rate A (in s$^{-1}$), population $n(i)$ and intensity $I_{ij}(\lambda)$ of E1 transition from 3s$^2$3p$^5$3d$^3$ to 3s$^2$3p$^6$3d$^2$ of W$^{54+}$ ion are presented for a wavelength range from 29.5-32.5 \AA\, in Table~\ref{Tab2}. The experimental value observed by EBIT and the theoretical values from FAC\cite{Gu2008The} and MCDF\cite{0953-4075-50-4-045004,Dipti2015Electron} are also included for comparison. The present calculation agrees quite well with the experimental data and other theoretical values. The wavelength discrepancy between T. Lennartsson \emph{et al.} \cite{PhysRevA.87.062505} by experiment, Dipti \emph{et al.} \cite{Dipti2015Electron} and Ding \emph{et al.} \cite{0953-4075-50-4-045004} by MCDF method, T. Lennartsson \emph{et al.} \cite{PhysRevA.87.062505} by FAC calculations and the present calculation are about 0.25\%, 0.14\%, 0.11\% and 0.07\%, respectively. All the observed transition lines from the EBIT experiment were identified in the present calculation. It was found that the observed transition lines in the EBIT experiment have large transition rates. However, a few transitions in this range with large transition rates have not been observed in the previous EBIT experiment, such as transitions with the key 7, 8 and 9. The reason is the population (Pop) of the excited upper levels of these unobserved transitions is extremely small. The results show that the intensity (Int) of the unobserved transitions are generally smaller by four orders of magnitude than the intensity which could be observed. The intensity might be changed with the plasma conditions. It can be expected that these unobserved strong transitions might be observed by the appropriate plasma condition.

\begin{table}
\caption{The calculated transition wavelength lambda (\AA), transition rate A(in $s^{-1}$), population (Pop), intensity (Int) and available experimental and other theoretical values for the strong E1 transition of W$^{54+}$ ion. The column 'Key' correspond to the label in figure~\ref{fig1}. Notion $a(b)$ for transition probabilities A means a$\times$10$^b$.}
\label{Tab2}
\scriptsize
\begin{tabular}{p{1.1cm}p{1.1cm}p{1.3cm}p{1.3cm}p{1.5cm}p{1.5cm}p{1.8cm}p{1.3cm}p{1.1cm}}\hline
 Lower	&	Upper   &	$\lambda(\AA)$	&	$\lambda_{others}(\AA)$		&	$A(/s^{-1}$)	& $A^{c}(/s^{-1}$)	&	Pop	&  Int  &Key    \\\hline		
1	&	30	&	32.403	&	32.264$^a$  32.502$^b$ 32.401$^c$ 32.416$^d$   	&	8.30(10)	&	8.50(10)	&	6.92(-14)	&	5.74(-3)&1\\	
1	&	31	&	31.787	&	31.811$^a$  31.783$^b$ 31.787$^c$ 31.786$^d$ 	&	7.49(11)	&	7.44(11)	&	3.63(-14)	&	2.72(-2)&2\\	
1	&	32	&	31.719	&	31.776$^a$ 31.765$^b$ 31.732$^c$ 31.711$^d$ 	&	5.43(11)	&	5.91(11)	&	9.05(-14)	&	4.92(-2)&3\\	
1	&	33	&	31.522	&	31.563$^a$ 31.503$^b$ 31.536$^c$ 31.505$^d$ 	&	9.60(11)	&	9.43(11)	&	6.35(-14)	&	6.09(-2)&4\\	
1	&	34	&	31.400	&	31.430$^a$ 31.378$^b$ 31.410$^c$ 31.386$^d$ 	&	5.89(11)	&	5.18(11)	&	8.80(-14)	&	5.18(-2)&5\\	
2	&	38	&	31.215	&	31.245$^a$ 31.236$^b$ 31.251$^c$ 31.155$^d$	    &	9.39(11)	&	9.33(11)	&	7.75(-15)	&	7.28(-3)&6\\	
6	&	68	&	31.077	&	31.115$^c$	                                    &	1.18(12)	&	1.16(12)    &   8.68(-20)	&	1.03(-7)&7\\	
7	&	79	&	30.924	&	30.948$^c$	                                    &	1.19(12)	&	1.17(12)    &	1.88(-20)	&	2.23(-8)&8\\	
5	&	69	&	30.866	&	30.898$^c$	                                    &	1.19(12)	&	1.16(12)    &	3.26(-16)	&	3.86(-4)&9\\	
1	&	38	&	29.489	&	29.560$^a$ 29.456$^b$ 29.530$^c$ 29.452$^d$ 	&	3.08(11)	&	2.96(11)	&	7.75(-15)	&	2.39(-3)&10\\
\hline
\end{tabular}
\begin{tablenotes}
\item[] $^{\rm a}$ From T. Lennartsson by EBIT experiment\cite{PhysRevA.87.062505}.
\item[] $^{\rm b}$ From Dipti et al by MCDF method\cite{Dipti2015Electron}.
\item[] $^{\rm c}$ From Ding et al by MCDF method\cite{0953-4075-50-4-045004}.
\item[] $^{\rm d}$ From T. Lennartsson by collisional-radiative
model\cite{PhysRevA.87.062505}\\
\end{tablenotes}
\end{table}

\begin{figure}
\centering
\includegraphics{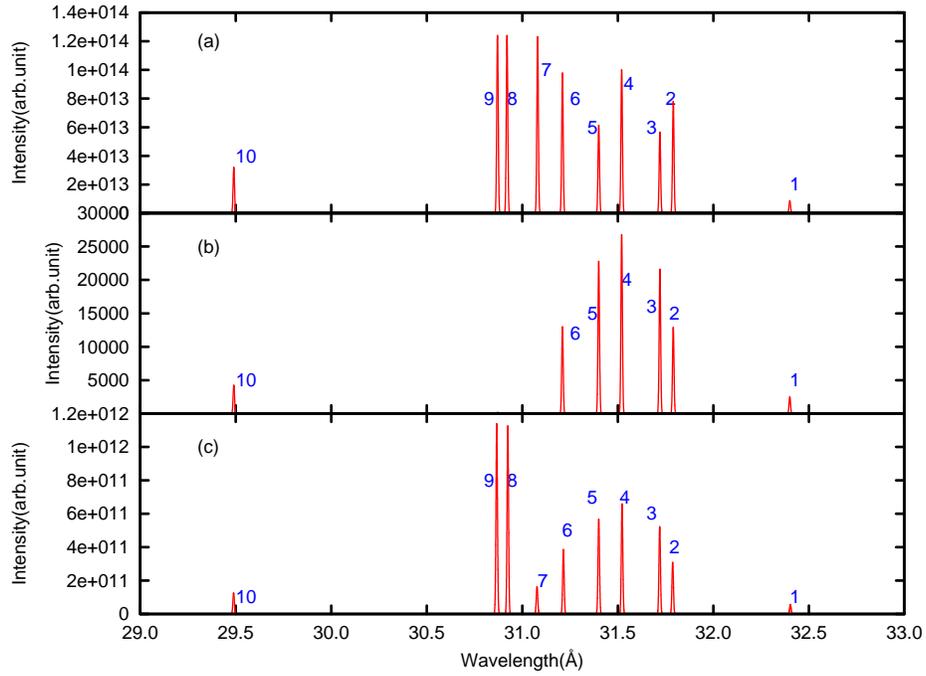}
\caption{The synthetic spectrum of W$^{54+}$ ion in wavelength 29.5-32.5 \AA. (a).convoluting the transition rate with the Gaussian profile; (b).the spectrum for the EBIT case with $n_e$ = 10$^{12}$ cm$^{-3}$ and $E_e$ = 18.2 keV; (c).the spectrum for the LTE plasma with $n_e$ = 10$^{15}$ cm$^{-3}$ and $T_e$ = 18.2 keV.}\label{fig1}
\end{figure}

The synthetic spectrum of W$^{54+}$ ion in the wavelength 29.5-32.5 \AA\, is shown in Fig.~\ref{fig1}. Each individual transition was assumed to have the Gaussian profile with full width at half maximum (FWHF) 0.09 \AA. The upper part Fig.~\ref{fig1}(a) is the spectrum obtained by convoluting the transition rate, while the middle part Fig.~\ref{fig1}(b) is the spectrum by considering the population of the upper level in the EBIT case with the electron density $n_e$ = 10$^{12}$ cm$^{-3}$ and the energy of electron beam $E_e$ = 18.2 keV. All 7 peaks observed by the experimental observation\cite{PhysRevA.87.062505} can be reproduced by the synthetic spectrum. The lower part Fig.~\ref{fig1}(c) is the spectrum by considering the excited upper levels in the LTE plasma with the electron density $n_e$ = 10$^{15} $ cm$^{-3}$ and the electron temperature $T_e$ = 18.2 keV, and the electron energy distribution is Maxwellian. The results indicate that all the strong transition lines are observable under this plasma condition.

The large difference between Fig.~\ref{fig1} (b) and (c) is caused by the dependence of population mechanism on the free electron energy distribution function in a different plasma environment. The intensity for a specific transition line is proportion to the population of the excited upper level $i$ which will be populated by the collisional excitation processes from other lower energy levels and collisional deexcitation processes from other higher energy levels (referenced as population flux), and will be depopulated by the collisional excitation processes to other higher energy levels and collisional deexcitation processes to other lower energy levels (referenced as depopulation flux). Collisional excitation and deexcitation rates coefficient are obtained by convoluting the cross section of the corresponding collisional (de)excitation processes with the free electron energy distribution function (EEDF) in the plasma. In the present work, the EEDF in the EBIT plasma and LTE plasma are taken as the $\delta$ function and the Maxwellian distribution function, respectively. In the EBIT plasma, for example, the strong transition line have been observed in the previous EBIT experiment with the key 4, the ratio of the population flux and depopulation flux of the excited upper level is 1.12, while the weak transition line which have not been observed in the previous EBIT experiment with the key 9, the ratio of the population flux and depopulation flux is 0.15. This means the excited upper level of the weak peak have a small population compare with the strong peak. In the LTE plasma, for all excited upper levels, the ratio of the population flux and depopulation flux about are 0.07, which means the population for the excited upper levels almost same. Thus the intensity only proportional to the transition rates. As a result, the transition 7, 8 and 9 are large enough to be observed.

\begin{figure}
\begin{center}
\includegraphics{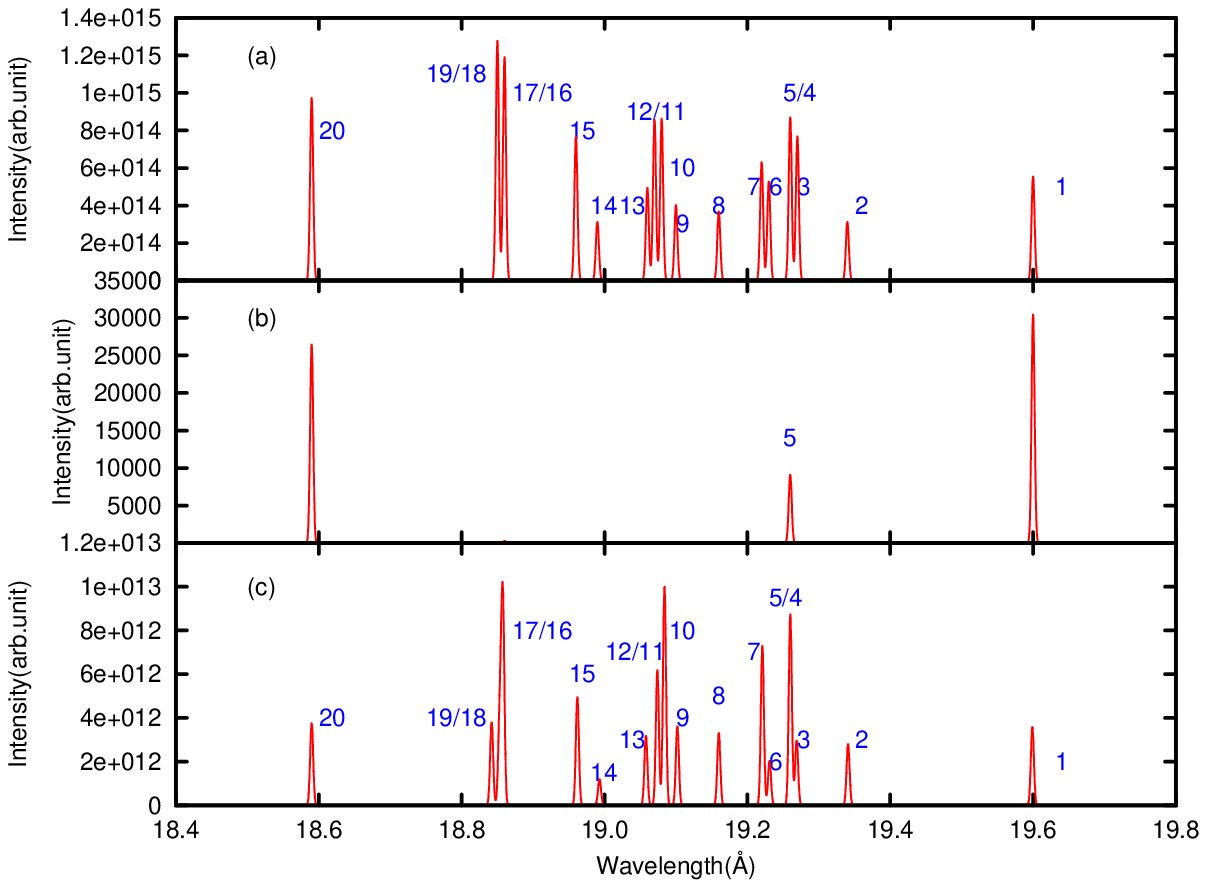}\\  % insert figure
\caption {The synthetic spectrum of W$^{54+}$ ion in wavelength 18.5-19.6 \AA. (a).convoluting the transition rate with the Gaussian profile; (b).the spectrum for the EBIT case with $n_e$ = 10$^{12}$ cm$^{-3}$ and $E_e$ = 18.2 keV; (c).the spectrum for the LTE plasma with $n_e$ = 10$^{15}$ cm$^{-3}$ and $T_e$ = 18.2 keV.}\label{Fig2}
\end{center}
\end{figure}

The wavelength $\lambda$ (in \AA), transition rate A (in s$^{-1}$), population $n(i)$ and intensity $I_{ij}(\lambda)$ of E1 transition 3s$^2$3p$^5$3d$^3$ to 3s$^2$3p$^6$3d$^2$ in the wavelength range 18.5-19.6 \AA\, are presented in Table~\ref{Tab3}. The theoretical values from MCDF calculation\cite{0953-4075-50-4-045004} are also included for comparison. The present calculation values generally made a good agreement with the previous data. The wavelength discrepancy was found to be about 0.04\%. According to the present calculation, some strong transitions may be observed in the wavelength range of 18.5-19.6 \AA. The synthetic spectrum of W$^{54+}$ ion in this wavelength range was shown in Fig.~\ref{Fig2}. All the peaks were obtained with FWHM = 0.05 \AA\, for each individually transition to make the spectrum clear. The upper part Fig.~\ref{Fig2}(a) is the spectrum obtained by convoluting the transition rate with the Gaussian profile. The middle part Fig.~\ref{Fig2}(b) is the spectrum by considering the population of the upper levels in the EBIT case with $n_e$ = 10$^{12}$ cm$^{-3}$ and $E_e$ = 18.2 keV. According to the synthetic spectrum there are only 3 peaks that would be observed in the EBIT experiment with this condition, namely, the transition lines with key 1, 5, and 20 in Fig.~\ref{Fig2}(b), These 3 peaks are regarded as the E1 transition {[(3p$_{1/2}$3p$_{3/2}^{4}$)$_{1/2}$(3d$_{3/2}^{3}$)$_{3/2}$]$_{2}$} $\rightarrow$ {[3p$^{6}$3d$_{3/2}^{2}$]$_{2}$}, {[(3p$_{1/2}$3p$_{3/2}^{4}$)$_{1/2}$(3d$_{3/2}^{3}$)$_{3/2}$]$_{1}$} $\rightarrow$ {[3p$^{6}$3d$_{3/2}^{2}$]$_{0}$}, {[(3p$_{1/2}$3p$_{3/2}^{4}$)$_{1/2}$(3d$_{3/2}^{3}$)$_{3/2}$]$_{1}$} $\rightarrow$ {[3p$^{6}$3d$_{3/2}^{2}$]$_{2}$} with transition wavelength 19.599 \AA, 19.260 \AA\, and 18.590 \AA, respectively. The lower part Fig.~\ref{Fig2}(c) is the spectrum by considering the population of the upper levels in the LTE plasma with $n_e$ = 10$^{15}$ cm$^{-3}$ and $T_e$ = 18.2 keV, and the electron energy distribution is Maxwellian. The results indicate that all transition lines are observable in this condition. The difference between Fig.~\ref{Fig2} (b) and (c) are caused by the same reason as in Fig.~\ref{fig1}.

\begin{table}
\caption{The calculated transition wavelength lambda (\AA), transition rate A(in $s^{-1}$), population (Pop), intensity (Int) and the theoretical values from MCDF calculation for the strong E1 transition of W$^{54+}$ ion. The column 'Key' correspond to the label in figure~\ref{Fig2}. Notion $a(b)$ for transition probabilities A means a$\times$10$^b$.}
\label{Tab3}
\scriptsize
\begin{tabular}{p{1.1cm}p{1.1cm}p{1.3cm}p{1.3cm}p{1.5cm}p{1.5cm}p{1.8cm}p{1.3cm}p{1.1cm}} \hline
Lower	&	Upper   &	$\lambda(\AA)$	&	$\lambda^{a}(\AA)$		&	$A(/s^{-1}$)	&	$A^{a}(/s^{-1})$	&	Pop	&  Int    &   Key    \\\hline		
1	&	82		&	19.599	&	19.603	&	2.95(12)	&	2.88(12)	&	1.31(-14)	&	3.85(-2)&1\\
8	&	109		&	19.341	&	19.340	&	1.66(12)	&	1.61(12)	&	1.32(-19)	&	2.19(-7)&2\\
8	&	110		&	19.269	&	19.266	&	4.09(12)	&	4.03(12)	&	2.58(-18)	&	1.06(-5)&3\\
7	&	107		&	19.261	&	19.264	&	2.85(12)	&	2.79(12)	&	9.40(-20)	&	2.68(-7)&4\\
2	&	83		&	19.260	&	19.261	&	1.78(12)	&	1.76(12)	&	4.86(-15)	&	8.64(-3)&5\\
9	&	113		&	19.231	&	19.232	&	2.81(12)	&	2.79(12)	&	1.84(-17)	&	5.15(-5)&6\\
7	&	108		&	19.221	&	19.220	&	3.36(12)	&	3.30(12)	&	5.78(-19)	&	1.94(-6)&7\\
4	&	91		&	19.160	&	19.162	&	1.96(12)	&	1.93(12)	&	3.71(-17)	&	7.26(-5)&8\\
7	&	109		&	19.102	&	19.102	&	2.14(12)	&	2.13(12)	&	1.32(-19)	&	2.82(-7)&9\\
5	&	92		&	19.084	&	19.086	&	4.59(12)	&	4.52(12)	&	2.44(-17)	&	1.12(-4)&10\\
8	&	111		&	19.074	&	19.084	&	1.46(12)	&	1.45(12)	&	6.80(-19)	&	9.90(-7)&11\\
8	&	112		&	19.074	&	19.078	&	3.10(12)	&	3.00(12)	&	1.06(-17)	&	3.29(-5)&12\\
6	&	93		&	19.058	&	19.060	&	2.63(12)	&	2.60(12)	&	2.11(-17)	&	5.55(-5)&13\\
6	&	94		&	18.993	&	18.991	&	1.66(12)	&	1.64(12)	&	1.66(-17)	&	2.74(-5)&14\\
3	&	90		&	18.962	&	18.967	&	4.08(12)	&	3.97(12)	&	4.48(-17)	&	1.83(-4)&15\\
5	&	95		&	18.858	&	18.865	&	4.35(12)	&	4.24(12)	&	1.86(-17)	&	8.11(-5)&16\\
3	&	91		&	18.856	&	18.861	&	1.99(12)	&	1.96(12)	&	3.71(-17)	&	7.40(-5)&17\\
4	&	94		&	18.853	&	18.852	&	4.55(12)	&	4.48(12)	&	1.66(-17)	&	7.55(-5)&18\\
7	&	111		&	18.842	&	18.852	&	2.25(12)	&	2.15(12)	&	6.80(-19)	&	1.53(-6)&19\\
1	&	83		&	18.590	&	18.594	&	5.18(12)	&	5.09(12)	&	4.86(-15)	&	2.52(-2)&20\\

\hline
\end{tabular}
\begin{tablenotes}
\item[] $^{\rm a}$From Ding et al by MCDF method\cite{0953-4075-50-4-045004}.
\end{tablenotes}
\end{table}

\section{Conclusion}

 The energy level, E1 transition rate, and electron collisional excitation of the ground state 3s$^2$3p$^6$3d$^2$ and the first excited state 3s$^2$3p$^5$3d$^3$ of W$^{54+}$ ion were calculated by relativistic configuration interaction method. A collisional-radiative model (CRM) was constructed to simulate the E1 transition spectrum for the EBIT and the LTE plasma. All the necessary atomic data for constructing the CRM was calculated by FAC packages. The most important configuration interaction effects were taken into account. The energy levels and transition rates made a reasonable agreement with the EBIT experimental observation and the previous theoretical values. The synthetic spectrum from the CRM explained the EBIT observation in the 29.5-32.5 \AA. Furthermore, some possible transitions were proposed to be observed in 18.5-19.6 \AA\, of the future the EBIT observations. Finally, the difference of the spectrum in different plasma condition was observed and explained.

\section*{Acknowledgment}

This work was supported by National Key Research and Development Program of China, Grant No:NYK0123, National Nature Science Foundation of China, Grant No: 11264035, Specialized Research Fund for the Doctoral Program of Higher Education (SRFDP), Grant No: 20126203120004, International Scientific and Technological Cooperative Project of Gansu Province of China (Grant No. 1104WCGA186), JSPS-NRF-NSFC A3 Foresight Program in the field of Plasma Physics (NSFC: No. 11261140328, NRF: 2012K2A2A6000443).

\section*{References}
\footnotesize
\bibliography{abc}
%\bibliography{mybibfile}

\end{document}